\documentclass[fleqn,twoside]{article}
\usepackage{espcrc2}

\usepackage{graphicx}
\usepackage[figuresright]{rotating}

\newcommand{\AmS}{{\protect\the\textfont2
  A\kern-.1667em\lower.5ex\hbox{M}\kern-.125emS}}

\title{APENet: LQCD clusters \`a la APE\thanks{Talk given at Lattice 2004 by R.A.}}

\author{R. Ammendola\address[INFNRM2]{Istituto Nazionale di Fisica Nucleare, Sezione Roma II \\
				      Via della Ricerca Scientifica 1, I-00133 Rome, Italy},
        M. Guagnelli\addressmark\address[UNIRM2]{Dipartimento di Fisica, Universit\`a di Roma {\em Tor Vergata}\\
						 Via della Ricerca Scientifica 1, I-00133 Rome, Italy},
        G. Mazza\addressmark[INFNRM2],
        F. Palombi\addressmark[INFNRM2]\address[FERMI]{{\em E.~Fermi} Research Center, c/o Compendio Viminale, pal.~F, I-00184 Rome, Italy},
	R. Petronzio\addressmark[INFNRM2]\addressmark[UNIRM2],
	D. Rossetti\address[INFNRM1]{Istituto Nazionale di Fisica Nucleare, Sezione Roma I \\
				     Piazzale Aldo Moro 2, I-00186 Rome, Italy},
	A. Salamon\addressmark[INFNRM2],
	and P. Vicini\addressmark[INFNRM1].}
       
\begin{document}

\begin{abstract}

Developed by the APE group, APENet is a new high speed, low latency,
3-dimensional interconnect architecture optimized for PC clusters
running LQCD-like numerical applications.  The hardware implementation
is based on a single PCI-X 133MHz network interface card hosting six
independent bi-directional channels with a peak bandwidth of 676 MB/s
each direction. We discuss preliminary benchmark results showing 
exciting performances similar or better than those found in high-end
commercial network systems.

\vspace{1pc}
\end{abstract}

\maketitle

\section{Overview}
The APE research group\cite{APEgroup} has traditionally focused on the
design and development of custom silicon, electronics and software
optimized for Lattice QCD (LQCD).

Recent works in LQCD numerical application area \cite{Luscher,Fodor,Lippert}
 have shown an increasing
interest on clusters of commodity PC's.  This is mainly due to two
facts: good sustained performance of commodity processors on numerical
applications and slowly emerging low latency, high bandwidth network
interconnects.

This paper describes APENet, a 3D network of point-to-point,
low-latency, high-bandwidth links well suited for medium sized
clusters running numerical applications.

\section{APENet}
APENet is a 3D network of point-to-point links with toroidal boundary
conditions.  Each Processing Element (PE), in our case a cluster node,
has 6 full-duplex communication channels ($X^+$, $X^-$, $Y^+$, $Y^-$,
$Z^+$, $Z^-$).

Data are transmitted in packets which are routed to the destination PE
according to simple --- and software overridable --- rules. Packet
delivery is always guaranteed: trasmission is delayed until the
receiver has enough room in its receive buffers.
No external routing device is necessary: \textit{next-neighbour}
and longer distance communications are obtained efficently hopping
until the destination PE is reached, without penalties for in-between
PEs.

Latency is kept to the minimum thanks to a lightweight low level
protocol --- just two 64bit words for the header and the footer, ---
and to the cut-through architecture of the switching device. Within 10
clock cycles from the arrival of the header, the receiving channel starts
forwarding the packet along its path, either toward local buffers ---
for packets intended for that very PE --- or toward the proper
trasmitting channel --- for packets which hop away. ---

\subsection{The APELink Card}
\begin{figure} [t] 
  \centering
  \includegraphics[width=\columnwidth,keepaspectratio]{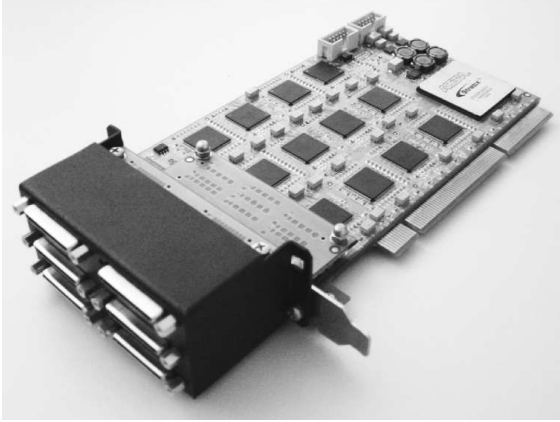}
  \vspace*{-10mm}
  \caption{\small The APELink card.}
  \label{fig:foto}
  \vspace*{-5mm}
\end{figure}

The building block of the APENet implementation is the APELink card,
shown in Fig.~\ref{fig:foto}
The APELink is a PCI-X 133MHz 64bit card which uses an Altera's
Stratix device, a last generation FPGA, as a the network device
controller, and six pairs of serializers/deserializers from
National Semiconductors as physical link interfaces.
\begin{figure} [bht]
  \centering
  \vspace*{-7mm}
  \includegraphics[angle=-90,width=\columnwidth,keepaspectratio]{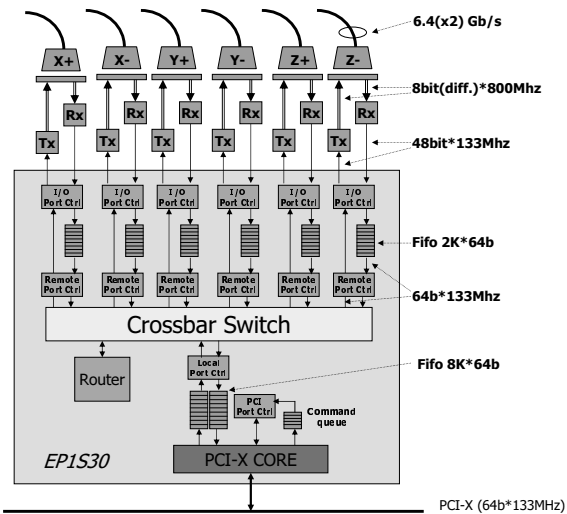}
  \vspace*{-10mm}
  \caption{\small The APELink functional blocks diagram.}
  \label{fig:apelink}
  \vspace*{-5mm}
\end{figure}
The APELink card, see Fig.~\ref{fig:apelink}, is composed of three
major functional blocks. Each block has its own clock domain and all
data communications between these blocks are based on dual clock
FIFOs, which guarantee robustness of the hardware itself. The first
block is the PCI-X interface, which handles the communication with the
host PCI-X bus; the second block, called \textit{crossbar switch},
controls the data flow among the PCI-X channel and the remote
communication \textit{links}; the third block implements six remote
communication bi-directional \textit{links}.

\subsection{The APENet Software}

The main programming interface is a proprietary simple
library (\texttt{apelib}) of C functions, including synchronous,
asyncronous and basic collective functions. It relies on the APELink
driver, a Linux device driver fully Multi Processor-aware (SMP) and
supporting versions 2.4 and 2.6 of the Linux kernel.  We are
developing both an MPI implementation based on LAM-MPI and a network
device driver, which allows simple IP protocol traffic to be routed on
the APENet.

The \texttt{apelib} is targeted for numerical application code and
includes basic primitives such as $\mathtt{ape\_send()}$,
$\mathtt{ape\_recv()}$, $\mathtt{ape\_sndrcv()}$, and some collective
functions ($\mathtt{ape\_broadcast()}$,
$\mathtt{ape\_global\_sum()}$). The $\mathtt{ape\_sndrcv()}$ primitive
squeezes the best performances from our architecture, as it
asymptotically exercises two channels at once, incrementing
the aggregated bandwidth.

\section{Benchmarks\label{sec:benchmarks}}

In this section we report some preliminary low-level benchmark
results, obtained on APELink early prototypes.

Benchmarks were performed on some dual Intel Xeon PC's, with both
ServerWorks GC-LE and Intel E7501 chipsets. The PC's are connected in
a small ring topology. The APELink channel speed is currently kept at
100 MHz with a peak performance of 508 MB/s per link while the PCI-X
interface runs at 133MHz.

The benchmark performs a "ping-pong" data transfer (unidirectional and
bi-directional) between two adjacent PE.  In the unidirectional test,
one PE sends a message to a remote PE then blocks on receiving a
response. The second PE receives the full message and sends back the
same amount of data. Half round-trip time, averaged on a number of
iterations, is defined as the latency, i.e. the message transfer time.

\begin{figure} [t]
  \centering
  \includegraphics[width=\columnwidth,keepaspectratio]{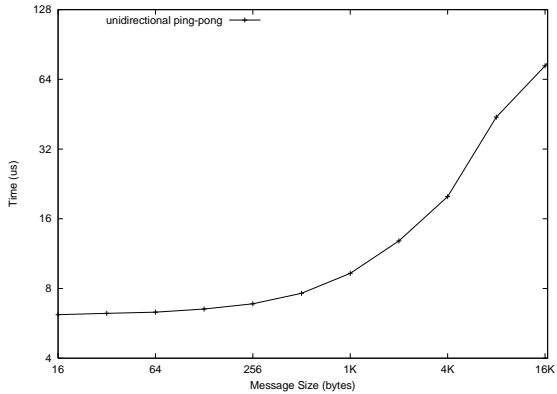}
  \vspace*{-10mm}
  \caption{\small Latency as measured in a ping-pong test for small packet sizes.}
  \label{Fig-Latency}
  \vspace*{-5mm}
\end{figure}

From the same test we have estimated the sustained bandwidth. The
bi-directional test differs from the unidirectional one since both PEs
send data simultaneously using the $\mathtt{ape\_sndrcv()}$ function.

\begin{figure} [htb]
  \centering
  \vspace*{-7mm}
  \includegraphics[width=\columnwidth,keepaspectratio]{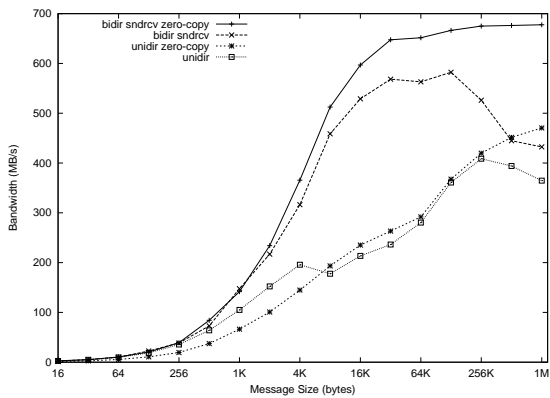}
  \vspace*{-10mm}
  \caption{\small Bandwidth is measured both for uni-directional and
  bidirectional tests. The \textit{zero-copy} label refers to the use
  of optimized memory buffers.}
  \label{Fig-Bandwidth}
  \vspace*{-5mm}
\end{figure}

In Fig.~\ref{Fig-Latency} we plot the latency for message sizes
ranging from 16 to 16K bytes. The smallest message size is 16 as the
minimum packet payload is a 128bit word. The estimated latency is
$\sim~6\mu s$ and is constant up to 256 bytes size message. For 4096
bytes messages we measure $20\mu s$ which is quite good and pretty
similar to commercial interconnects\cite{benchIBA}.

Fig.~\ref{Fig-Bandwidth} shows the bandwidth plot with
message sizes ranging from 16 bytes to 1MB.  The bi-directional
zero-copy bandwidth saturates at 677 MB/s. At 1MB message size the
uni-directional bandwidth is 470MB/s, roughly $90\%$ of the channel peak
performance at 100MHz.  The plot shows two pairs of curves: those marked
\textit{zero-copy} refer to the use of pinned-down memory, suitable to
be used for PCI DMA transfers. This way the overhead of expensive
memory copy operations to/from DMA memory buffers are avoided. Non
\textit{zero-copy} data are reported only to simplify the discussion.
 
\section{Conclusions}

The hardware design of the APElink card is completed and we are
running tests on the final release of the board whose link channels
run at full speed (133MHz).  Preliminary benchmarks have shown
encouraging results, comparable with commercial network
interconnects. The APELink software is currently in fast progress:
current activities focus on a better low level driver 
a MPI implementation.

The INFN prototype APENet PC cluster, composed of 16 PC's equipped with
APElink boards, is ready to be used on LQCD test codes.  We have
plans to expand it up to 64 PC's ($4^3$ topology) in the near future.


\begin{thebibliography}{9}

\bibitem{APEgroup}
The APE group, Istituto Nazionale di Fisica Nucleare\\
\texttt{http://apegate.roma1.infn.it/APE}


\bibitem{Luscher}
M.~Luscher,
Nucl.\ Phys.\ Proc.\ Suppl.\  {\bf 106}, 21 (2002)
[arXiv:hep-lat/0110007].

\bibitem{Fodor}
Z.~Fodor and S.~D.~Katz,
JHEP {\bf 0203}, 014 (2002)
[arXiv:hep-lat/0106002].

\bibitem{Lippert}
T.~Lippert,
Nucl.\ Phys.\ Proc.\ Suppl.\  {\bf 129}, 88 (2004)
[arXiv:hep-lat/0311011].  



\bibitem{benchIBA}
J.~Liu {\it et al.},
\textit{Performace Comparison of MPI Implementations over Infiniband, Myrinet and Quadrics},
SuperComputing Conference, November 2003.




\end{thebibliography}
\end{document}